\documentclass[3p,preprint,times]{elsarticle}

\usepackage{ecrc}

\volume{489C}

\firstpage{18}

\journalname{Physica A}

\runauth{M. Sekania et al.}

\jid{physa}

\jnltitlelogo{Physica A}

\usepackage{amsmath}
\usepackage{amssymb}
\usepackage{graphicx}
\usepackage{bm}
\usepackage{color}

\usepackage{lineno,hyperref}
\modulolinenumbers[5]

\journal{Physica A}
\begin{document}

\begin{frontmatter}

\title{Scaling behavior of the Compton profile of alkali metals}

\author[a,b]{Michael Sekania\corref{corrauth}}
\address[a]{Theoretical Physics III, Center for Electronic Correlations and Magnetism, Institute of Physics, University of Augsburg, D-86135 Augsburg, Germany}
\address[b]{Andronikashvili Institute of Physics, Tamarashvili 6, 0177 Tbilisi, Georgia}
\cortext[corrauth]{Corresponding author}
\ead{msekania@physik.uni-augsburg.de}

\author[a,c]{Wilhelm H. Appelt}
\address[c]{Augsburg Center for Innovative Technologies, University of Augsburg, D-86135 Augsburg, Germany}

\author[d,e]{Diana Benea}
\address[d]{Faculty of Physics, Babes-Bolyai University, Str. M. Kogalniceanu nr. 1, Ro-400084 Cluj-Napoca, Romania}
\address[e]{Department of Chemistry, University Munich, Butenandstr.~5-13, D-81377 M\"unchen, Germany}

\author[e]{Hubert Ebert}

\author[a]{Dieter Vollhardt}

\author[a,c]{Liviu Chioncel}

\begin{abstract}
The contribution of the valence electrons to the Compton profiles of the crystalline alkali
metals is calculated using density functional theory. We show that the
Compton profiles can be modeled by a $q-$Gaussian distribution, which is
characterized by an anisotropic, element dependent parameter $q$.
Thereby we derive an unexpected scaling behavior of the Compton profiles of all
alkali metals.
\end{abstract}


\end{frontmatter}

\section{Introduction}
The Compton effect, discovered in 1923, provides a basic illustration
of the conservation of energy and momentum in quantum mechanical processes \cite{comp.23}.
Since then the inelastic scattering of an X-ray photon from a charged particle, usually an electron in a target, with a large energy and momentum transfer, is referred to as \emph{Compton scattering}.
The inevitable quantum mechanical motion of the target electrons leads to a Doppler shift of the Compton scattered photon and thereby to a broadening of the Compton lineshape. This so-called \emph{Compton profile} furnishes important information about the momentum distribution of the target electrons
(for reviews see \cite{coop.85,co.mi.04}).
Indeed, it was a great success and a turning point in the early history of the quantum theory of solids when in 1929
DuMond~\cite{mond.29,mond.30,mond.33}
measured the Compton profile of beryllium and thereby showed that electrons obey Fermi-Dirac statistics, which had been proposed only three years earlier.

In a Compton scattering experiment
the double differential scattering cross section
$d^2 {\sigma}/d \Omega d\omega$ is measured.
Within the \emph{impulse approximation}~\cite{ch.wi.52,cu.de.71}
the latter is directly proportional to the Compton profile $J(p_z)$, which is an integral over the momentum density $n({\bf p})$
in the plane perpendicular to the scattering vector $\bf{K}$:
\begin{equation}\label{cp1}
\frac{d^2 \sigma}{d \Omega d\omega} \propto J(p_z)=
\iint
n({\bf p}) dp_x dp_y; \quad (p_z || {\bf K}).
\end{equation}
The momentum density in a quantum system
is defined as the average number of particles with momentum {\bf p}:
$n({\bf p}) = \langle \Psi | \sum_{\sigma} a_{{\bf p} \sigma}^\dagger
a_{{\bf p} \sigma}  | \Psi \rangle $.
Here the normalized $N$-particle state of the system is represented by
$| \Psi \rangle $
and $a_{{\bf p} \sigma}^\dagger (a_{{\bf p} \sigma})$
are the creation (annihilation) operators for particles with momentum {\bf p}
and spin projection $\sigma$.
The single-particle momentum density $n({\bf p})$ plays a significant
role in our understanding of ground state properties of quantum many-particle systems.
Outstanding candidates are interacting Fermi systems~\cite{land.57,land.57.bis,nozi.thin}
such as liquid $^3$He, electrons in metals, and atomic nuclei.

There are a  few cases in which the shape of the Compton
profile can be determined exactly \cite{coop.85}.
For instance,
the Compton profile for a non-interacting electron gas is
simply an inverted parabola for momenta $p<p_F$, where $p_F$ is the
Fermi momentum, and zero otherwise: $ J(p) \propto (p_F^2 -p^2) \theta(p_F-p)$.
For an isolated atom, or in the case of scattering with one bound
state~\cite{kr.kr.77},
the Compton profile takes the form of
a Lorentzian.
In all other cases $J(p)$ develops a non-trivial tail for large momenta.

In solids with a periodic lattice potential continuous translational invariance is broken
and the momentum ${\bf p}$ is no longer proportional to the wave vector ${\bf k}$ introduced by
Bloch's theorem \cite{ashcroft}.
The single particle momentum density is then given by \cite{coop.85,co.mi.04,schu.78}
\begin{equation}\label{nofp_explicit}
  n(\mathbf{p}) = \sum_{b, \mathbf{k}, {\bf G}} |u_{b,{\bf k+G}}|^2
  n_{bb}(\mathbf{k}) \delta (\mathbf{k}+ {\bf G} - \mathbf{p}),
\end{equation}
where $\bf{G}$ is a translation vector in the reciprocal lattice,
$n_{bb}(\mathbf{k})$ is the orbital-resolved {\em occupation density},
which has a diagonal representation in the {\em natural orbitals} of L\"owdin \cite{lowd.51},
and $u_{b,{\bf k+G}}$ are the Fourier components of the natural orbitals.
In contrast to the non-interacting electron gas,
the Compton profile of the electrons in solids continues beyond the Fermi momentum $p_F$.
It exhibits infinitely many cusps with diminishing amplitude in the case of
conduction electrons of alkali metals,
while the core electrons generally lead to a very broad and smooth Compton profile.
By varying the energy of the incident photon in
a Compton scattering experiment it is possible to distinguish between the contributions of the core and conduction electrons~\cite{pl.tz.65}.
The Compton profiles measured for different orientations of the probe
are used to reconstruct the momentum density,
providing information about Fermi surface features
and the directional anisotropy due to the underlying crystal structure \cite{coop.85,co.mi.04,ba.zo.99,kont.09,dugd.14}.
Its Fourier transform, which is connected to the so called reciprocal form factor \cite{ashcroft}, is a well-studied quantity,
containing information about the chemical bonds in the crystal structure \cite{coop.85,co.mi.04,sc.st.96}.

In this paper we compute the Compton profiles of the conduction electrons of the crystalline elemental metals of the first column of the periodic table --- the alkali metals
Li, Na, K, Rb, Cs --- within the local-density approximation (LDA) of density functional theory (DFT)~\cite{jo.gu.89,kohn.99,jone.15}.
Based on our DFT(LDA) calculations we will show that the
global shape of the Compton profile can be fitted by a
{\it $q-$}Gaussian distribution \cite{tsal.88}.
Thereby we demonstrate a previously unnoticed scaling behavior of the
Compton profiles for \emph{all} alkali metals, which allows us to
collapse all data points. This scaling  suggests that the multiple scattering of
valence (mobile) electrons in solids
can be viewed as a stochastic dynamics that asymptotically produces the generalized canonical distribution introduced by Tsallis~\cite{tsal.88}.

\section{Compton profiles of alkali metals}

\subsection{Low momenta range}
A considerable number of theoretical and experimental studies of Compton scattering have
been carried out for Li (see Refs.~\cite{ei.la.72,sa.ta.95,sc.st.96,kubo.97,schl.99,ba.zo.99,fi.ce.99,st.ha.00,ta.sa.01,st.bu.01,sc.st.01,bros.02,bros.05}, and references therein).
However, none of the experiments report measurement results above $4$ atomic units ($a.u.$) of momentum.
Fewer studies have been reported for Na and K, and even less for Rb and
Cs~\cite{ei.la.72,tan.73,sob.85,coop.85,hu.ha.01,st.ha.00,co.mi.04,ol.ti.12}.
In all these studies, spectra were analyzed in a momentum range up to $2~a.u.$, but the tail behavior at larger momenta was never determined.

The first high resolution measurements of CP of Li, supplemented by numerical results, were reported by Sakurai {\it et al.}~\cite{sa.ta.95} and Sch\"ulke {\it et al.}~\cite{sc.st.96}.
The overall shapes of the measured CP as well as their first and second derivatives were found to be similar to the theoretical predictions (in the studied region).
However, at low momenta the theoretical results overestimate, and at higher momenta ($p_z \gtrsim p_F$) underestimate, the experimental values of CP.
Partially contradictory theoretical interpretations were reported (for details see \cite{bros.05} or the discussion part of \cite{sc.st.01}).
The discrepancies between the theoretical and the experimental results have been mainly attributed to the insufficient treatment of electron-correlations within the LDA framework.
Several schemes such as the isotropic Lam-Platzman corrections~\cite{la.pl.74} based on the electron momentum distribution of the interacting homogeneous electron gas have been employed with different success.
Sakurai {\it et al.}~\cite{sa.ta.95} showed that these corrections reduce the discrepancy between theory and experiment.
However the overall effect was found to be relatively small.
Related studies by Tanaka {\it et al.}~\cite{ta.sa.01} came to the similar conclusions.
QMC studies of Li~\cite{fi.ce.99} do not reveal any substantial difference in the electron momentum density with respect to the LDA predictions.
The perturbative GW calculations by Kubo~\cite{kubo.97}, which include correlation effects beyond density functional theory, seemed to be in best agreement with the experimental CPs.
However, Sch\"ulke pointed out in Ref.~\cite{schl.99} that these 
results contained some numerical instabilities. 
Later DFT studies involving more sophisticated functionals~\cite{ca.co.89,fa.he.93} 
describing the electron momentum density of a Fermi liquid found a sufficient 
agreement between experimental and theoretical results~\cite{ba.zo.99,sc.st.01,bros.05}.
Recently Aguiar~\cite{ag.da.15} proposed yet another semi-empirical parametrization of the electron momentum density of the Fermi liquid
in order to describe the electron momentum densities and CPs for metals.
This parametrization, however, does not take the directional anisotropy into account.

The discrepancies between conventional band theory and experiment might also originate from the processing of experimental data, which include
systematic errors, incorrect subtraction of (possibly non-linear) background, multiple scattering or core electron contribution, or incorrect estimation of the resolution function.
Bross~\cite{bros.05} pointed out the appearance of a dip at $p_z=0$ in the reconstructed electron momentum density from the experimental results Ref.~\cite{ta.sa.01},
and raised doubts about the accuracy of the experimental CPs for small values of $p_z$.
In order to reproduce the experimental results up to few $p_F$-s, it is most likely that one has to take several effects into account including finite temperature, lattice expansions, thermal disorder, correlation corrections.

At present it is widely accepted that the LDA overestimates the CP at lower momenta ($p < p_F$) and underestimates the large momenta tails ($p \gtrsim p_F$) (see e.g. \cite{co.mi.04}).
Proposed LP corrections \cite{ba.zo.99,sc.st.01,bros.05} distinctly differ up to $(2-3)p_F$, but almost coincide beyond this range.
In addition, all these corrections become negligibly small as compared to uncorrected LDA results in this outer region.

\begin{figure}[t]
\begin{center}
   \includegraphics[width=0.7\textwidth, clip=true]{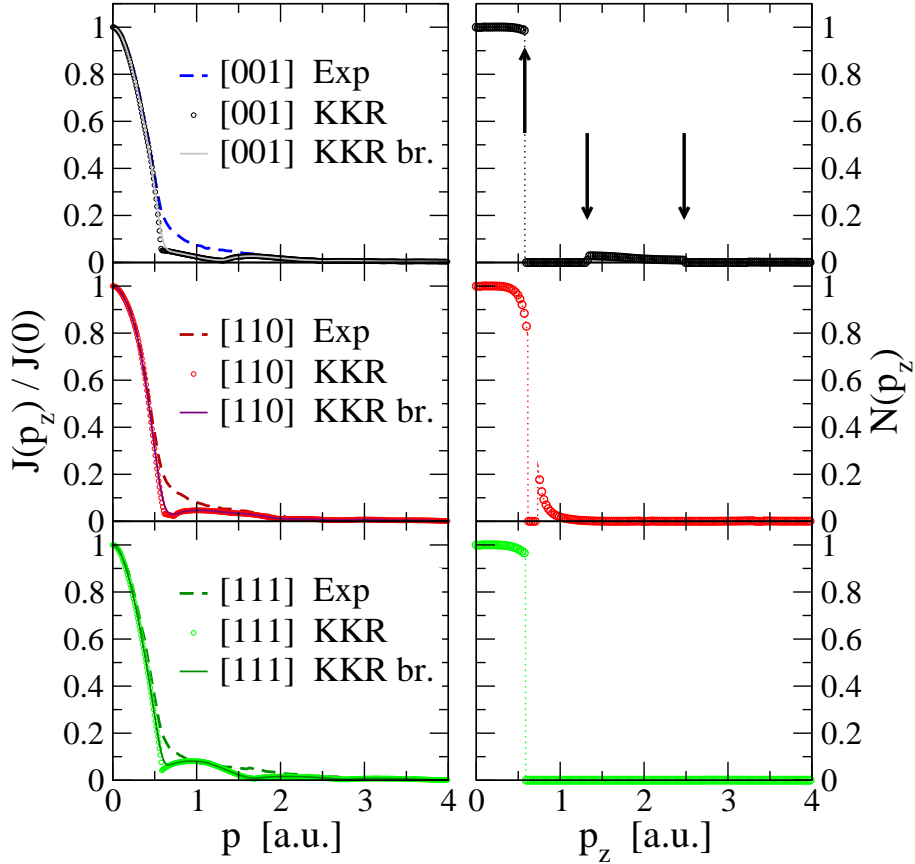}
\end{center}
\caption {\label{Fig:figure1}
Compton profiles (left column) and
momentum distributions (right column) of
Li calculated along the principal directions. Fermi-surface
contributions at higher momenta are visible along the [001] and [110] directions and are indicate by arrows.
Theoretical data convoluted with the Gaussian with full width at half maximum of $0.12\,a.u.$, corresponding to the experimental resolution, is shown with solid lines (KKR br.).
Experimental results without core contribution (dashed lines in the left column) are taken from Ref.~\cite{sa.ta.95}.
}
\end{figure}

\subsection{Moderate and high momenta range}
Although high resolution is achievable in modern experiments \cite{sa.ta.95,sc.st.96,ta.sa.01,sc.st.01},
the measurements of CP for momenta higher than a few $a.u.$ was not reported up to now.
Measurements for larger momenta (up to $10~a.u.$) have been performed \cite{ta.sa.01}, but unfortunately the CP of valence electrons were reported only up to $3~a.u$..
Other studies were also targeting the first Brillouin zone and CP up to a few $p_F$-s.
Nevertheless the Compton profile at higher momenta contains important physical information since it originates from the tails of the momentum distribution caused by many-body effects and the lattice periodicity.
Already a decade ago, Bross drew attention to the importance of momenta $p \gg p_F$ for the accurate calculation of the Compton profile for Li \cite{bros.05}.
Analytic calculations of the tails of the momentum distribution are usually based on many-body perturbation theory.
For example, for the interacting electron gas (without lattice) the tails fall off as $p^{-8}$ in lowest order perturbation
theory~\cite{da.vo.60,ya.ka.76}. Recently, exact relations for the jellium model  were derived
using the operator product expansion technique~\cite{ho.ba.13},
and power law tails were identified in two and three dimensions.
For metallic densities the interaction cannot be treated by perturbation methods and therefore
the tails need to be computed numerically, e.g., using
quantum Monte Carlo techniques~\cite{dr.ne.09,ho.be.11}.
In this case, the exchange-correlation functional is the crucial quantity determining the momentum density and the shape
of the Compton profile, including tails.

\begin{figure}[!t]
\begin{center}
   \includegraphics[width=0.7\textwidth, clip=true]{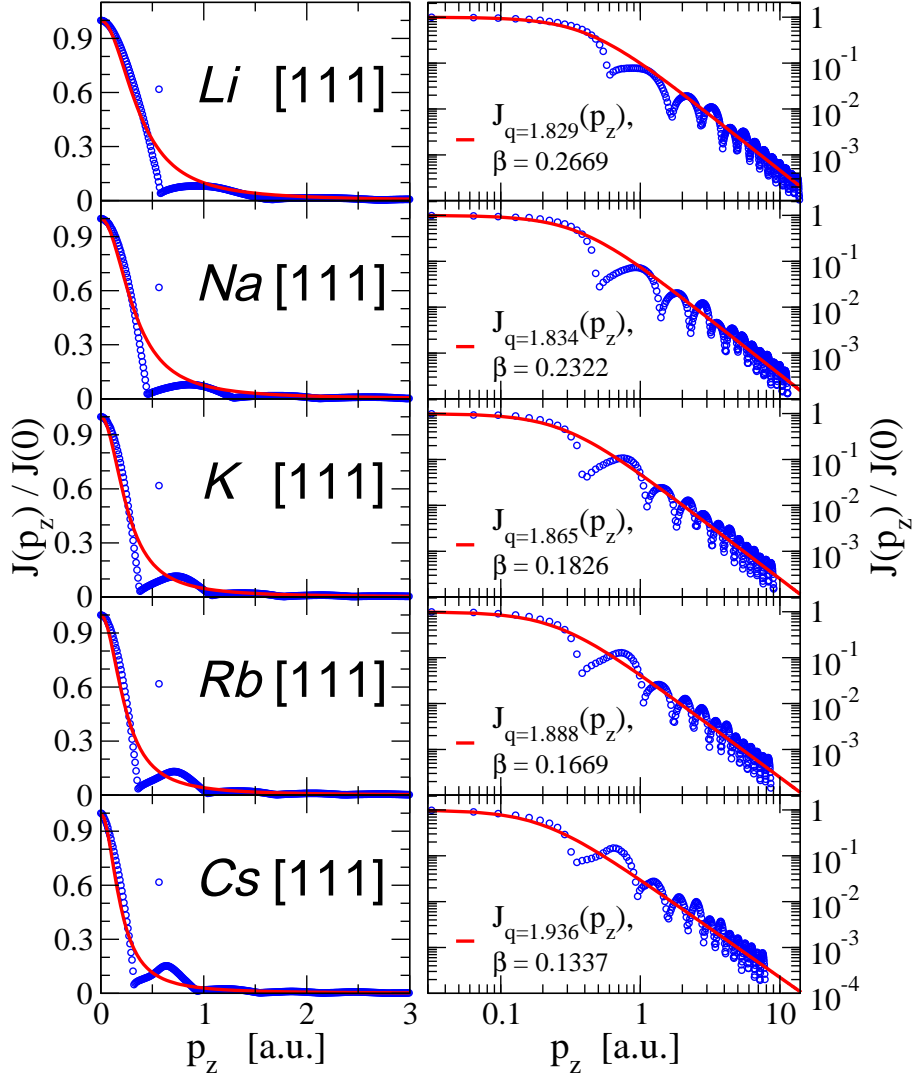}
\end{center}
\caption {\label{Fig:figure3} Compton profiles of the
alkali metals Li, Na, K, Rb and Cs calculated along the principal direction $[111]$.
$q$-Gaussian fit to the lineshape (red solid line) normalized to $J(p_z = 0)=1.0$;
$\beta$ is given in atomic units ($a.u.$).}
\end{figure}

In our study the electronic structure of alkali metals was calculated
within DFT \cite{jo.gu.89,kohn.99,jone.15} using the spin-polarized relativistic
Korringa-Kohn-Rostoker (SPR-KKR) method
~\cite{eb.ko.11}, which was recently extended
to compute Compton and magnetic Compton profiles (MCPs)~\cite{sz.gy.84,be.ma.06}.
The exchange-correlation potentials parametrized by
Vosko, Wilk and Nusair~\cite{VWN80} were employed for calculations in the local spin-density approximation (LSDA).
For the integration over the Brillouin zone the special points method was
employed~\cite{mo.pa.76}.
The spin resolved momentum densities were
computed from the corresponding LSDA Green functions in momentum space as
\begin{equation}\label{e7}
n_{m_s}(\vec p)={-\frac{1}{\pi} \Im \int_{-\infty}^{E_F}
G_{m_s}(\vec p,\vec p,E)dE},
\end{equation}
where $m_s=\uparrow,\downarrow$.
The electron momentum densities are usually calculated for the three
principal directions $[001]$, $[110]$, $[111]$ using a rectangular grid
of about thousand points in each direction. In our calculation the maximum value of the momentum is $16~a.u.$ in each direction.
Compton profiles are either normalized by the area under the curve (equal to the number of valence electrons), or by the intensity at zero momentum
$J(p_z=0)=1$.

In Fig.~\ref{Fig:figure1} we show the computed Compton profiles of the single-conduction-band electrons of solid BCC Li along the three principal directions in comparison with experiment \cite{sa.ta.95}.
To compare with experiment a broadening of the computed results is required.
The computed core contributions are subtracted from the experimental data \cite{sa.ta.95}.
When compared with experimental values, the Compton profiles computed by us support the widely known behavior, namely, they overestimate at low momenta and underestimate at higher momenta the experimentally measured values~\cite{sa.ta.95,sc.st.96,ba.zo.99,fi.ce.99}.
The Compton profiles have a parabola-like shape for $p_z<p_F$ (first cusp)
and pronounced tails for $p_z>p_F$.
Higher momentum contributions to $n({\bf p})$  \cite{Schuelke.1977,sc.st.96,ba.zo.99,bros.05} are clearly seen along [110] and [001] directions and are evidence for Umklapp processes.
Hence, the anisotropy in the momentum distribution is mainly a consequence of the directional anisotropy of the bcc lattice.

In the left column of Fig.~\ref{Fig:figure3} we show the Compton
profiles (Eq.~\ref{cp1}) of all alkali metals computed along the direction [111]
of the bcc structure as a function of the momentum component $p_z$.
The first cusp occurs at the Fermi momenta located at
$p_F^{Li} = 0.58\,a.u.$, $p_F^{Na} = 0.49\,a.u.$,
$p_F^{K} = 0.39\,a.u.$, $p_F^{Rb} = 0.36\,a.u.$
and $p_F^{Cs} = 0.32\,a.u.$.
At $p_z=0$ the relation
$J^{Li}(0)< J^{Na}(0)<J^{K}(0)<J^{Rb}(0)<J^{Cs}(0)$ holds (this is not visible in Fig.~\ref{Fig:figure3} due to the normalization by the intensity at zero momentum).
Further cusps follow at higher momenta.
The intensity of the Compton profiles at $p_z=10\,p_F$ is about three orders of magnitude lower
than that at $p_z=0$.

\section{$q$-Gaussian modeling}
We will now show that the overall shape of the Compton profile of the conduction electrons of the alkali metals, i.e., the parabola-like behavior at low momenta and the algebraic tails at high momenta, is well described by a $q-$Gaussian distribution~\cite{tsal.88},
which belongs to the family of leptocurtic distributions.
The $q$-Gaussian probability distribution~\cite{tsal.88} is defined as
\begin{equation}\label{eq:JqG}
J_q(p_z)=\frac{1}{\sqrt{2}\,\beta\,C_q}\exp_q(-\frac{p_z^2}{2\beta^2}),
\end{equation}
where $exp_q$ is the $q$-analog of the exponential function
\begin{equation}
\exp_q(x)=(1+(1-q)x)^{1/(1-q)}
\end{equation}
and $C_q$ is a normalization factor.
This distribution reduces to an inverted parabola in the limit $q\rightarrow 0$ and to the Gaussian distribution for $q\rightarrow 1$. In particular, it exhibits algebraic tails, $J_q(p_z)\sim 1/p_z^{2/(q-1)}$, for $1<q<3$ at large values of $p_z$.

\begin{figure}[t]
\begin{center}
   \includegraphics[width=0.5\textwidth, clip=true]{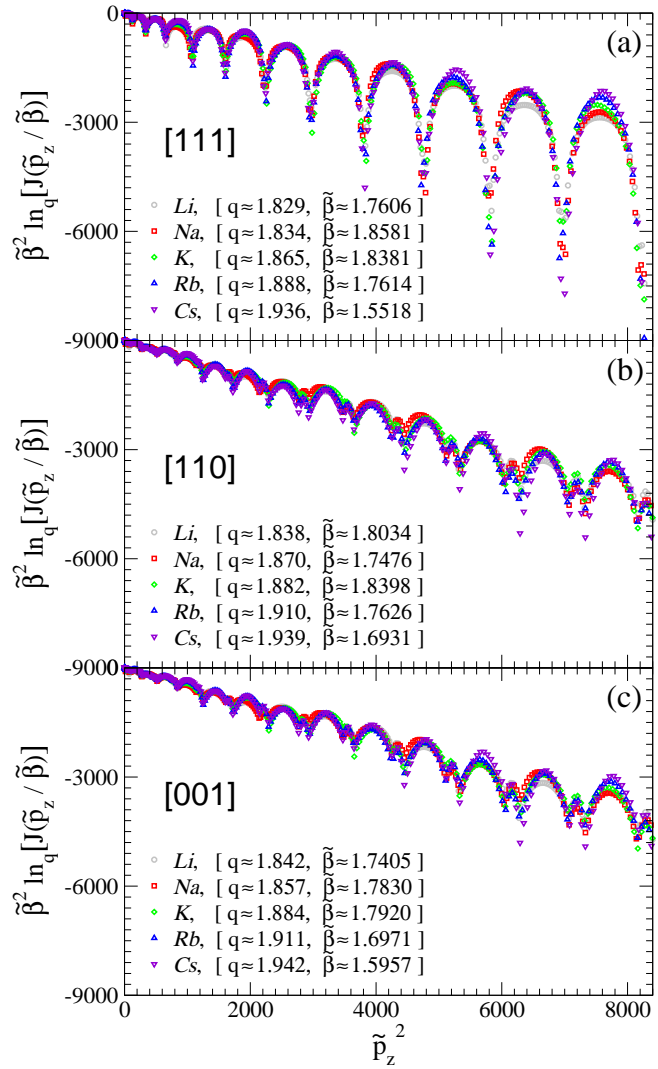}
\end{center}
\caption {\label{Fig:figure9} Rescaled Compton profiles
vs. $\tilde{p}_z^2=(p_z\,a_0)^2$ for different scattering directions ((a): [111];
(b): [110]; (c): [001]); see text. The element specific $(q,\tilde{\beta})$
parameters are seen to be anisotropic.}
\end{figure}

Assuming the problem under investigation is fully isotropic, we can estimate the asymptotic behavior of the valence-electron momentum density.
For the isotropic case, $n({\mathbf p})=n(p)$ is directly obtained from the isotropic CP \cite{bros.05,mijn.77}: 
\begin{equation}
n(p)=-\frac{1}{2\pi p}\frac{\mathrm{d}J(p)}{\mathrm{d}p}, \qquad \mathrm{for} \quad p\neq 0\,.
\end{equation}
Considering a $q$-Gaussian as an approximation of the CP lineshape, $J(p)=J_q(p)$,
we find $n(p) \sim 1/p^{2q/(q-1)}$, for $1<q<3$ at large values of $p_z$, for the electron momentum density.

The algebraic tails of the Compton profiles are superposed by cusps which are a straightforward consequence of the underlying lattice periodicity.
By fitting the overall shape of the Compton profile (i.e., without the cusps) by a $q$-Gaussian distribution, Eq.~\eqref{eq:JqG}, both the
deviation from the parabola-like form at low momenta and the asymptotic behavior of the tails at large momenta can be analyzed, and the values of the ``entropic'' parameter $q$  and the spread $\beta$ can be extracted (see right panel of Fig.~\ref{Fig:figure3}).
Going from the lighter to heavier elements in the first column,
relativistic effects are found to be less than $1\%$ for Li and Na,
approximately $1\%$ for K and $8\%$ for Rb, more than $10 \%$ for Cs~\cite{schw.04}.
This is to be expected since elements with larger nuclear charge $Z$ are subject to stronger relativistic
effects, resulting in an increasing relativistic mass and a decreasing orbital radius (inversely
proportional to the mass).
In solids, orbitals of valence electrons, subjected to the lattice potential,
form the bands which are explicitly captured within DFT~\cite{jo.gu.89,kohn.99,jone.15,eb.ko.11}.
As the fits to the $q$-Gaussian function reveal, the valence band electrons for heavier elements approach the limit of core electrons
with larger values of the entropic parameter ($q\rightarrow 2$).
At the same time the spread $\beta$ decreases.

\section{Scaling analysis}
The fact that the overall shape of the Compton profiles can be fitted by $q$-Gaussian distributions, Eq.~\eqref{eq:JqG}, allows one to perform a scaling analysis of the Compton profiles in terms of the inverse relation
\begin{equation}
 {\tilde{\beta}}^{2} \ln_q\left[J\left(\frac{\tilde{p}_z}{\tilde{\beta}} \right)\right] = -\frac{1}{2}{\tilde{p}_z}^{2}.
\end{equation}
Here $ ln_q$ is the corresponding $q$-analog of the logarithm defined by
\begin{equation}\label{eq:log_q}
    \ln_q x :=  \frac{x^{1-q}-1}{1-q}\,
\end{equation}
and $\tilde{p}_z = p_z a_0$ and $\tilde{\beta} = \beta a_0$ are the quantities $p_z$ and $\beta$, respectively, expressed in dimensionless units.
The scaling plots are shown in
Fig.~\ref{Fig:figure9} for the three principal direction of the crystal, where
$ {\tilde{\beta}}^{2} \ln_q J(\tilde{p}_z/\tilde{\beta})$ is plotted against ${\tilde{p}_z}^{2}$.
The parameters $q$ are taken
from the $q$-Gaussian fits to the
corresponding Compton profile.
The scaling of the Compton profiles of the alkali metals, which is found to hold up to the highest values of the momentum component $p_z$, is very remarkable and unexpected. The $(q,\beta)$ parameters differ along the three principal directions.

\section{Summary and Discussion}
In summary, the Compton profiles of the single-conduction-band of crystalline alkali metals were found to deviate significantly from the free-electron (inverted parabola) form below $p<p_F$ and to develop algebraic tails for high momenta.
In particular, we showed that the overall shape of the Compton profiles can be fitted by $q-$Gaussian distributions from moderate to high momenta. This led us to derive a scaling relation which allows one to collapse the Compton profiles of all alkali metals along a given principal direction of the lattice onto a single curve.

In view of the fact that the $q-$Gaussian distribution is an exact stationary solution of the standard linear Fokker-Planck equation \cite{borl.98} and was recently found to describe the stationary momentum distribution of cold atoms in dissipative optical lattices~\cite{lutz.03,lutz.04,lutz.13}, our results suggest that the multiple scattering of valence electrons in solids
may be understood as a stochastic dynamics which asymptotically produces the $q-$Gaussian distribution.
In contrast to cold atoms in optical lattices where the potential can be tuned to produce different entropic parameters $q$, the electrons in solids are subject to the potential determined by the
underlying lattice structure and chemical composition.
Hence the $q-$Gaussian distribution has a fixed entropic parameter $q$.
In a general Kohn-Sham construction the multiple scattering equations reduce to the solution of the one-electron equation in which the effective one-electron potential is a functional of the density of the electrons in the system.
In the language of scattering theory the one-electron equation describes the collision of the electron with the external potential, representing electron-ion and electron-electron Coulomb interactions.
Alkali metals provide a unique possibility to study the relation between the electron-ion interaction strength and the corresponding value of $q$.
In fact, Vignat {\em et al.}~\cite{vi.pl.12} recently pointed out that in quantum mechanics the ground-state wave function of a particle in a Coulomb potential has the form of a $q$-Gaussian in momentum space.
The $q$-Gaussian in Ref.~\cite{vi.pl.12} is the square root of the $q$-Gaussian in our investigation, which is consistent with the fact that the momentum density is the square of the wave function in momentum representation.
Whether our findings reflect a specific shape of the screened Coulomb potential in \emph{real} solids,
captured by the special form of the exchange correlation function, remains a subject of further theoretical and experimental investigations.
Although CP values at the higher momentum are less accurate due to a poor counting statistics, our analysis showed robustness of the fitting procedure against noise.
Therefore it should be possible to verify our findings experimentally.
This will open the possibility to develop models where the entropic parameter is explicitly included in the multiple scattering formalism,
accounting for the important systematics in the data.

\section{Acknowledgments}
We thank C.~Tsallis for bringing Ref. \cite{vi.pl.12} to our attention.
We gratefully acknowledge financial support by the COST Action MP 1306, the
Deutsche Forschungsgemeinschaft through the Research Unit FOR 1346, the DAAD, and the
CNCS - UEFISCDI (project number PN-II-RU-TE-2014-4-0009 HEUSPIN).

\appendix
\section{$q$-Gaussian distribution function}
\label{qgauss}
\begin{figure}[t]
\begin{center}
   \includegraphics[width=0.57\textwidth, clip=true]{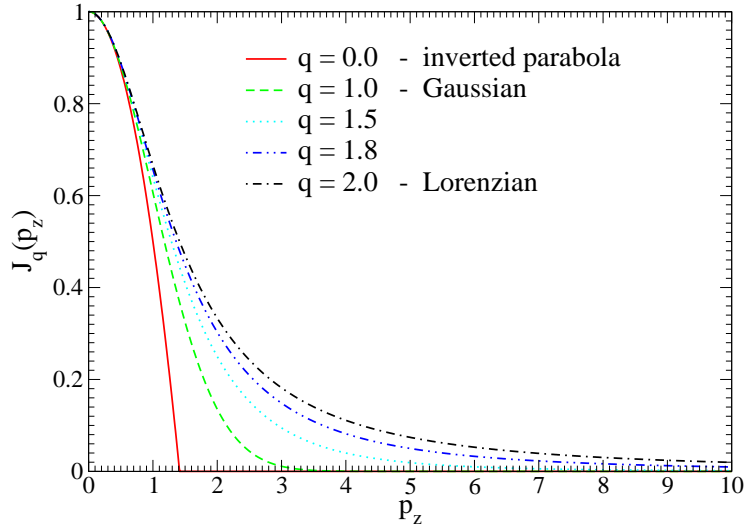}
\end{center}
\caption {\label{Fig:figure_app_1} Shape of the $q$-Gaussian distribution
for different values of $q$ at $\beta=1$.
}
\end{figure}

In our analysis of the scaling properties of the Compton profiles of the alkali metals we employ a generalization of the Gaussian distribution,
the so called $q$-Gaussian \eqref{eq:JqG}.
As discussed in the main text,
the family of $q$-Gaussians reproduces the Compton profile
in two well-known limiting cases: (i) For $q=0$ the $q$-Gaussian
has the shape of an inverted parabola, corresponding to the free, non-interacting
electron gas \cite{coop.85}; the curve reaches zero at $p_z=p_F$. (ii)
For $q=2$ the $q$-Gaussian reduces to a Lorentzian which describes the
Compton profile of bound scatterers, such as the core electrons in solids~\cite{kr.kr.77}.
Fig.~\ref{Fig:figure_app_1} shows the $q$-Gaussians \eqref{eq:JqG}
for different values of $q$ at $\beta=1$ on a linear scale.

\begin{figure}[t]
\begin{center}
   \includegraphics[width=0.52\textwidth, clip=true]{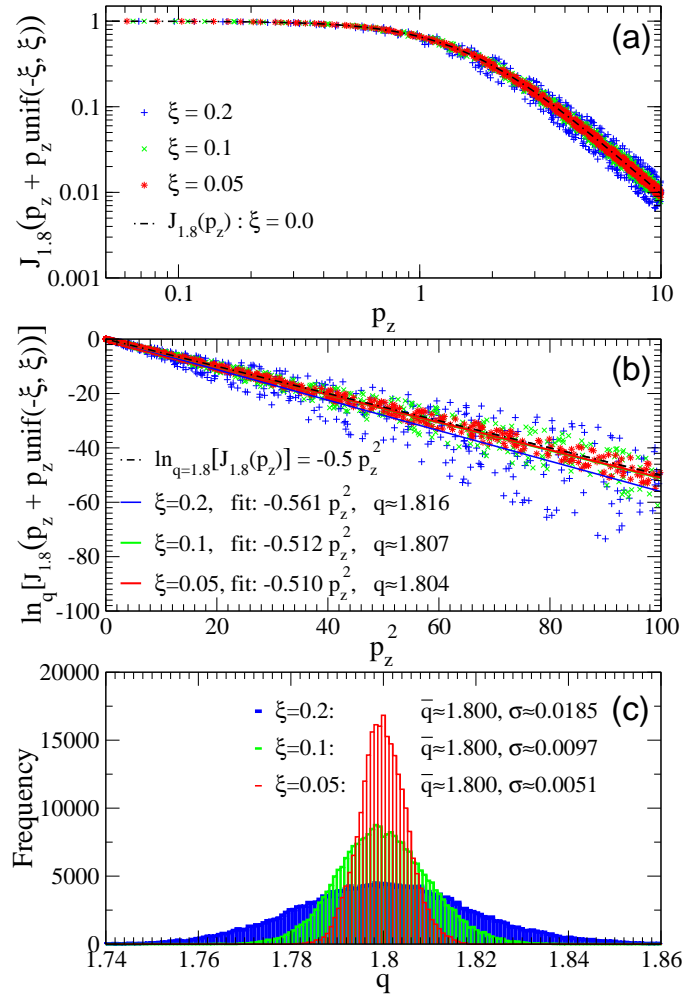}
\end{center}
\caption {\label{Fig:figure_app_2} (a),(b): q-Gaussian (black dash-dot line) and q-Gaussian samples with
$5\%$ ($\xi=0.05$, red ${*}$), $10\%$ ($\xi=0.1$, green $\times$), and $20\%$ ($\xi=0.2$, blue +)
multiplicative white noise, on a (a) linear and (b) double logarithmic scale, both for $q=1.8$ and $\beta=1$.
In panel (b) the best q-Gaussian fits are also shown.
(c): Histograms of the distribution of the best $q$-Gaussian fits for the $2.5\times 10^5$ random samples
of each set with $\xi=0.05$ (red), $\xi=0.1$ (green), and $\xi=0.2$ (blue).}
\end{figure}

For an accurate fit of the computed Compton profiles with a $q$-Gaussian distribution one has to take into account
that the numerical values of the Compton profile decrease in accuracy for
large values of $p_z$.
A least-square fit of the calculated data on a linear scale
would overemphasize the values of the Compton profile for small values of $p_z$,
since the intensity of the profile is large in this range.
On the other hand, a least-square fit on a double logarithmic scale
would overemphasize the values of the Compton profile for large values of $p_z$, i.e., in
the less accurate tail.
In order to fit the entire Compton profile lineshape
we require a fitting procedure which is able to treat the function at small and large $p_z$ on equal footing.
The difficulty in obtaining the optimal values of $q$ and $\beta$ lies mainly  in the accurate determination of $q$.
Therefore we separate the problem into two parts. First, assuming that the optimal value of $q$ is already known,
we take the inverse of the $q$-exponential, namely the $q$-logarithm \eqref{eq:log_q},
and apply this function to the obtained data and the $q$-Gaussian:
\begin{equation}\label{eq:log_q_q_gaussian}
   \ln_q (J_q(p_z)) = \ln_q(J_q(0)) - \left[1+(1-q)\ln_q J_q(0)\right]\left(\frac{p_z}{2\beta}\right)^2.
\end{equation}
By normalizing the initial data, ${J_q(p_z=0)} = 1$, one obtains
\begin{equation}\label{eq:log_q_q_gaussian_norm}
   \ln_q (J_q(p_z)) = -\left(\frac{p_z}{2\beta}\right)^2.
\end{equation}
For each $q$ value this defines a linear regression problem of the data points $(\ln_q (J_q(p_z)),{p_z}^2)$
and the global fitting procedure may be viewed as a one-dimensional optimization problem for the $q$ values.
Second, a criterion has to be found which allows one to determine the optimal value of $q$.
The squared deviation, $\chi^2$, of the linear fit of $\ln_q(J(p_z))$ vs. $p_z^2$, which is the typical measure for the fit accuracy in the linear regression problem,
is unsuitable here since $\chi^2$ is scaled down for large $q$ values (the data is flattened with growing $q$).
Hence, an optimization procedure for $q$ based on $\chi^2$ would always be biased towards large values of $q$.
A better alternative is suggested by examining Eq.~\eqref{eq:log_q_q_gaussian_norm}. Namely,
for the optimal value of $q$ the linear fit of $\ln_q(J(p_z))$ vs. $p_z^2$ should yield a line through the origin.
It turns out that there is only one value of $q$ which satisfies this criterion, at least for $q\in[1,3]$.
The slope $\alpha$ of the best least square fit then yields $\beta$ as
\begin{equation}\label{eg:fitte_beta}
 \beta = -\frac{2}{\alpha}.
\end{equation}

The stability of the  fitting procedure outlined above has been verified by the following procedure.
We take a $q$-Gaussian distribution for given values of $q$ and $\beta$ and introduce
 random, multiplicative white noise with amplitude $\xi$ in the argument of the function as
$p_z \rightarrow p_z + p_z\mathrm{unif}(-\xi,\xi)$, where $\mathrm{unif}(-\xi,\xi)$ indicates a uniform distribution.
We consider three different amplitudes: $\xi=0.05, 0.1, 0.2$, corresponding to $5\%$, $10\%$, and $20\%$ noise, respectively, and produce $2.5 \times 10^5$ samples for each $\xi$, which we refer to as ``set''.
By refitting the produced data we obtain the statistics on the reliability of the fit.
In Fig.~\ref{Fig:figure_app_2} we show the $q$-Gaussian for $q=1.8$, $\beta = 1.0$
and a single sample from each data set.
The choice of this type of noise is motivated by the realistic
envelope of error-bars around the data for the calculated Compton profile.
In Fig.~\ref{Fig:figure_app_2} we also show the best $q$-Gaussian fits for the presented samples
as well as the histograms which demonstrate the distribution of the best $q$-Gaussian fits for the $2.5\times 10^5$ random samples for each of the data sets.
As one can see the average value of $q$ of these distributions corresponds to $\bar{q}\approx1.8$. The standard deviations (data spread)
decrease for decreasing noise amplitudes.
These results validate the stability of the above mentioned fitting procedure.

\section*{References}

\biboptions{sort&compress}

\begin{thebibliography}{10}
\expandafter\ifx\csname url\endcsname\relax
  \def\url#1{\texttt{#1}}\fi
\expandafter\ifx\csname urlprefix\endcsname\relax\def\urlprefix{URL }\fi
\expandafter\ifx\csname href\endcsname\relax
  \def\href#1#2{#2} \def\path#1{#1}\fi

\bibitem{comp.23}
A.~H. {C}ompton, A quantum theory of the scattering of {X}-rays by light
  elements, Phys. Rev. 21 (1923) 483--502.

\bibitem{coop.85}
M.~J. Cooper, {C}ompton scattering and electron momentum determination, Reports
  on Progress in Physics 48~(4) (1985) 415.

\bibitem{co.mi.04}
M.~J. Cooper, P.~E. Mijnarends, N.~Shiotani, N.~Sakai, A.~Bansil, {X}-ray
  {C}ompton scattering, Oxford University Press, Oxford, 2004.

\bibitem{mond.29}
J.~W.~M. Du~Mond, {C}ompton modified line structure and its relation to the
  electron theory of solid bodies, Phys. Rev. 33 (1929) 643--658.

\bibitem{mond.30}
J.~W.~M. DuMond, Breadth of {C}ompton modified line, Phys. Rev. 36 (1930)
  146--147.

\bibitem{mond.33}
J.~W.~M. Dumond, The linear momenta of electrons in atoms and in solid bodies
  as revealed by {X}-ray scattering, Rev. Mod. Phys. 5 (1933) 1--33.

\bibitem{ch.wi.52}
G.~F. Chew, G.~C. Wick, The impulse approximation, Phys. Rev. 85 (1952)
  636--642.

\bibitem{cu.de.71}
R.~Currat, P.~D. DeCicco, R.~J. Weiss, Impulse approximation in {C}ompton
  scattering, Phys. Rev. B 4 (1971) 4256--4261.

\bibitem{land.57}
L.~D. Landau, Sov. Phys. JETP 3 (1957) 920.

\bibitem{land.57.bis}
L.~D. Landau, Sov. Phys. JETP 5 (1957) 101.

\bibitem{nozi.thin}
P.~Nozi{\`e}res, Theory of Interacting Fermi Systems, Bejamin, New York, 1964.

\bibitem{kr.kr.77}
B.~Kramer, P.~Krusius, Structure dependence of {C}ompton profiles. model study,
  Phys. Rev. B 16 (1977) 5341--5349.

\bibitem{ashcroft}
N.~W. Ashcroft, N.~D. Mermin, Solid State Physics, Saunders College,
  Philadelphia, 1976.

\bibitem{schu.78}
W.~Sch\"ulke, Solid state information from the {F}ourier transform of {C}ompton profiles,
  \href{http://stacks.iop.org/1347-4065/17/i=S2/a=332}{Japanese Journal of Applied Physics 17~(S2) (1978) 332}.

\bibitem{lowd.51}
P.-O. L\"{o}wdin, A note on the quantum-mechanical perturbation theory, J.
  Chem. Phys. 19~(11) (1951) 1396--1401.

\bibitem{pl.tz.65}
P.~M. Platzman, N.~Tzoar, {X}-ray scattering from an electron gas, Phys. Rev.
  139 (1965) A410--A413.

\bibitem{ba.zo.99}
T.~Baruah, R.~R. Zope, A.~Kshirsagar,
  Full-potential {LAPW} calculation of electron momentum density and related properties of {L}i,
  \href{http://link.aps.org/doi/10.1103/PhysRevB.60.10770}{Phys. Rev. B 60 (1999) 10770--10775}.
\newblock \href {http://dx.doi.org/10.1103/PhysRevB.60.10770}{\path{doi:10.1103/PhysRevB.60.10770}}.

\bibitem{kont.09}
G.~Kontrym-Sznajd,
  Fermiology via the electron momentum distribution ({R}eview {A}rticle),
  \href{http://scitation.aip.org/content/aip/journal/ltp/35/8/10.1063/1.3224712}{Low Temperature Physics 35~(8) (2009) 599--609}.
\newblock \href{http://dx.doi.org/10.1063/1.3224712}
  {\path{doi:10.1063/1.3224712}}.

\bibitem{dugd.14}
S.~B. Dugdale,
  Probing the {F}ermi surface by positron annihilation and {C}ompton scattering,
  \href{http://scitation.aip.org/content/aip/journal/ltp/40/4/10.1063/1.4869588}{Low Temperature Physics 40~(4) (2014) 328--338}.
\newblock \href{http://dx.doi.org/10.1063/1.4869588}{\path{doi:10.1063/1.4869588}}.

\bibitem{sc.st.96}
W.~Sch\"ulke, G.~Stutz, F.~Wohlert, A.~Kaprolat, Electron momentum-space
  densities of {L}i metal: A high-resolution {C}ompton-scattering study, Phys.
  Rev. B 54 (1996) 14381--14395.

\bibitem{jo.gu.89}
R.~O. Jones, O.~Gunnarsson,
  The density functional formalism, its applications and prospects,
  \href{http://link.aps.org/doi/10.1103/RevModPhys.61.689}{Rev. Mod. Phys. 61 (1989) 689--746}.
\newblock \href {http://dx.doi.org/10.1103/RevModPhys.61.689}
  {\path{doi:10.1103/RevModPhys.61.689}}.

\bibitem{kohn.99}
W.~Kohn, Nobel lecture: Electronic structure of matter-wave functions and
  density functionals, Rev. Mod. Phys. 71 (1999) 1253--1266.

\bibitem{jone.15}
R.~O. Jones, Density functional theory: {I}ts origins, rise to prominence, and future,
\href{http://link.aps.org/doi/10.1103/RevModPhys.87.897}{Rev. Mod. Phys. 87 (2015) 897--923}.
\newblock \href {http://dx.doi.org/10.1103/RevModPhys.87.897}
  {\path{doi:10.1103/RevModPhys.87.897}}.

\bibitem{tsal.88}
C.~Tsallis, Possible generalization of {B}oltzmann-{G}ibbs statistics, Journal
  of Statistical Physics 52~(1-2) (1988) 479--487.

\bibitem{ei.la.72}
P.~Eisenberger, L.~Lam, P.~M. Platzman, P.~Schmidt, {X}-ray {C}ompton profiles
  of {L}i and {N}a: Theory and experiments, Phys. Rev. B 6 (1972) 3671--3681.

\bibitem{sa.ta.95}
Y.~Sakurai, Y.~Tanaka, A.~Bansil, S.~Kaprzyk, A.~T. Stewart, Y.~Nagashima,
  T.~Hyodo, S.~Nanao, H.~Kawata, N.~Shiotani, High-resolution {C}ompton
  scattering study of {L}i: {A}sphericity of the {F}ermi surface and electron
  correlation effects, Phys. Rev. Lett. 74 (1995) 2252--2255.

\bibitem{kubo.97}
Y.~Kubo, Effects of electron correlations on {C}ompton profiles of {L}i and {N}a in the {GW}
  approximation, \href{http://dx.doi.org/10.1143/JPSJ.66.2236}{J. Phys. Soc. Jpn. 66~(8) (1996) 2236--2239}.
\newblock \href {http://arxiv.org/abs/http://dx.doi.org/10.1143/JPSJ.66.2236}
  {\path{arXiv:http://dx.doi.org/10.1143/JPSJ.66.2236}}, \href{http://dx.doi.org/10.1143/JPSJ.66.2236} {\path{doi:10.1143/JPSJ.66.2236}}.

\bibitem{schl.99}
W.~Sch\"ulke, Comment on "{E}ffects of {E}lectron {C}orrelation..." by {Y}. {K}ubo,
  {J}.~{P}hys.~{S}oc.~{J}pn.~{\bf 66}~(1997)~2236, \href{http://dx.doi.org/10.1143/JPSJ.68.2470}{Journal of the Physical
  Society of Japan 68~(7) (1999) 2470--2471}.
\newblock \href {http://arxiv.org/abs/http://dx.doi.org/10.1143/JPSJ.68.2470}
  {\path{arXiv:http://dx.doi.org/10.1143/JPSJ.68.2470}}.

\bibitem{fi.ce.99}
C.~Filippi, D.~M. Ceperley, Quantum {M}onte {C}arlo calculation of {C}ompton
  profiles of solid lithium, Phys. Rev. B 59 (1999) 7907--7916.

\bibitem{st.ha.00}
C.~Sternemann, K.~H\"am\"al\"ainen, A.~Kaprolat, A.~Soininen, G.~D\"oring,
  C.-C. Kao, S.~Manninen, W.~Sch\"ulke, Final-state interaction in {C}ompton
  scattering from electron liquids, Phys. Rev. B 62 (2000) R7687--R7690.

\bibitem{ta.sa.01}
Y.~Tanaka, Y.~Sakurai, A.~T. Stewart, N.~Shiotani, P.~E. Mijnarends,
  S.~Kaprzyk, A.~Bansil,
  Reconstructed three-dimensional electron momentum density in lithium: {A} {C}ompton scattering study,
  \href{http://link.aps.org/doi/10.1103/PhysRevB.63.045120}{Phys. Rev. B 63 (2001) 045120}.
\newblock \href{http://dx.doi.org/10.1103/PhysRevB.63.045120}
  {\path{doi:10.1103/PhysRevB.63.045120}}.

\bibitem{st.bu.01}
C.~Sternemann, T.~Buslaps, A.~Shukla, P.~Suortti, G.~D\"oring, W.~Sch\"ulke,
  Temperature influence on the valence {C}ompton profiles of aluminum and lithium,
  \href{http://link.aps.org/doi/10.1103/PhysRevB.63.094301}{Phys.Rev. B 63 (2001) 094301}.
\newblock\href{http://dx.doi.org/10.1103/PhysRevB.63.094301}
  {\path{doi:10.1103/PhysRevB.63.094301}}.

\bibitem{sc.st.01}
W.~Sch\"ulke, C.~Sternemann, A.~Kaprolat, G.~D\"oring,
  Ultra-high resolution {C}ompton scattering of {L}i metal: Evaluation with respect to the
  correlation corrected occupation number density,
  \href{//www.degruyter.com/view/j/zpch.2001.215.issue-11/zpch.2001.215.11.1353/zpch.2001.215.11.1353.xml}{Zeitschrift f\"ur Physikalische Chemie 215  1353}.
\newblock \href{http://dx.doi.org/10.1524/zpch.2001.215.11.1353}
  {\path{doi:10.1524/zpch.2001.215.11.1353}}.

\bibitem{bros.02}
H.~Bross, Investigation of some ground state properties of lithium with the all
  electron {MAPW} method, physica status solidi (b) 229~(3) (2002) 1359--1370.

\bibitem{bros.05}
H.~Bross, Electronic structure of {L}i with emphasis on the momentum density and the {C}ompton profile,
  \href{http://link.aps.org/doi/10.1103/PhysRevB.72.115109}{Phys. Rev. B 72 (2005) 115109}.
\newblock \href{http://dx.doi.org/10.1103/PhysRevB.72.115109}
  {\path{doi:10.1103/PhysRevB.72.115109}}.

\bibitem{tan.73}
B.~W. Tan, The influence of electron-electron correlation and crystal structure
  on the {C}ompton profiles of lithium, sodium and potassium, J. Phys. F.:
  Metal Physics 3 (1973) 1716.

\bibitem{sob.85}
M.~Sob, Electron momentum density and the momentum density of positron annihilation pairs in alkali
  metals: high-momentum components, \href{http://stacks.iop.org/0305-4608/15/i=8/a=008}{Journal of Physics F: Metal Physics 15~(8) (1985) 1685}.

\bibitem{hu.ha.01}
S.~Huotari, K.~H\"am\"al\"ainen, S.~Manninen, A.~Issolah, M.~Marangolo,
  Assymtery of {C}ompton profiles, Journal of Physics and Chemistry of Solids
  62 (2001) 2205--2213.

\bibitem{ol.ti.12}
V.~Olevano, A.~Titov, M.~Ladisa, K.~H\"am\"al\"ainen, S.~Huotari, M.~Holzmann,
  Momentum distribution and {C}ompton profile by the \textit{ab initio} {GW}
  approximation, \href{http://link.aps.org/doi/10.1103/PhysRevB.86.195123}{Phys. Rev. B 86 (2012) 195123}.
\newblock \href{http://dx.doi.org/10.1103/PhysRevB.86.195123}
  {\path{doi:10.1103/PhysRevB.86.195123}}.

\bibitem{la.pl.74}
L.~Lam, P.~M. Platzman, Momentum density and {C}ompton profile of the
  inhomogeneous interacting electronic system. {I}. {F}ormalism, Phys. Rev. B
  9~(12) (1974) 5122--5127.
\newblock \href{http://dx.doi.org/10.1103/PhysRevB.9.5122}
  {\path{doi:10.1103/PhysRevB.9.5122}}.

\bibitem{ca.co.89}
D.~A. Cardwell, M.~J. Cooper,
  The effect of exchange and correlation on the agreement between {APW} and {LCAO} {C}ompton profiles
  and experiment, \href{http://stacks.iop.org/0953-8984/1/i=47/a=007}{Journal of Physics: Condensed Matter 1~(47) (1989) 9357}.

\bibitem{fa.he.93}
B.~Farid, V.~Heine, G.~E. Engel, I.~J. Robertson,
  Extremal properties of the {H}arris-{F}oulkes functional and an improved screening calculation
  for the electron gas, \href{http://link.aps.org/doi/10.1103/PhysRevB.48.11602}{Phys. Rev. B 48 (1993) 11602--11621}.
\newblock \href{http://dx.doi.org/10.1103/PhysRevB.48.11602}
  {\path{doi:10.1103/PhysRevB.48.11602}}.

\bibitem{ag.da.15}
J.~C. Aguiar, D.~Mitnik, H.~O.~D. Rocco,
  Electron momentum density and {C}ompton profile by a semi-empirical approach,
  \href{http://www.sciencedirect.com/science/article/pii/S0022369715000785}{Journal of Physics and Chemistry of Solids 83 (2015) 64 -- 69}.
\newblock \href{http://dx.doi.org/http://dx.doi.org/10.1016/j.jpcs.2015.03.023}
  {\path{doi:http://dx.doi.org/10.1016/j.jpcs.2015.03.023}}.

\bibitem{da.vo.60}
E.~Daniel, S.~H. Vosko, Momentum distribution of an interacting electron gas,
  Phys. Rev. 120 (1960) 2041--2044.

\bibitem{ya.ka.76}
H.~Yasuhara, Y.~Kawazoe,
  A note on the momentum distribution function for an electron gas,
  \href{http://www.sciencedirect.com/science/article/pii/0378437176900601}{Physica A: Statistical Mechanics and its Applications 85~(2) (1976) 416 -- 424}.
\newblock \href{http://dx.doi.org/http://dx.doi.org/10.1016/0378-4371(76)90060-1}
  {\path{doi:http://dx.doi.org/10.1016/0378-4371(76)90060-1}}.

\bibitem{ho.ba.13}
J.~Hofmann, M.~Barth, W.~Zwerger, Short-distance properties of {C}oulomb
  systems, Phys. Rev. B 87 (2013) 235125.

\bibitem{dr.ne.09}
N.~D. Drummond, R.~J. Needs, Quantum {M}onte {C}arlo study of the ground state
  of the two-dimensional {F}ermi fluid, Phys. Rev. B 79 (2009) 085414.

\bibitem{ho.be.11}
M.~Holzmann, B.~Bernu, C.~Pierleoni, J.~McMinis, D.~M. Ceperley, V.~Olevano,
  L.~Delle~Site, Momentum distribution of the homogeneous electron gas, Phys.
  Rev. Lett. 107 (2011) 110402.

\bibitem{eb.ko.11}
H.~Ebert, D.~K\"odderitzsch, J.~Min\'ar, Calculating condensed matter
  properties using the {KKR}-green's function method¿recent developments and
  applications, Reports on Progress in Physics 74~(9) (2011) 096501.

\bibitem{sz.gy.84}
Z.~Szotek, B.~L. Gyorffy, G.~M. Stocks, W.~M. Temmerman, Electron and
  electron-positron momentum distributions in concentrated random alloys, J.
  Phys. F.: Metal Physics 14~(11) (1984) 2571.

\bibitem{be.ma.06}
D.~Benea, S.~Mankovsky, H.~Ebert, Fully relativistic description of magnetic
  {C}ompton profiles with an application to
  $\mathrm{U}{\mathrm{\uppercase{f}e}}_{2}$, Phys. Rev. B 73 (2006) 094411.

\bibitem{VWN80}
S.~H. Vosko, L.~Wilk, M.~Nusair, Canadian Journal of Physics 58 (1980) 1200.

\bibitem{mo.pa.76}
H.~J. Monkhorst, J.~D. Pack, Special points for {B}rillouin-zone integrations,
  Phys. Rev. B 13 (1976) 5188--5192.

\bibitem{Schuelke.1977}
W.~Sch\"ulke, Fermi surface mapping from the one-dimensional {F}ourier transform of {C}ompton profiles,
  \href{http://dx.doi.org/10.1002/pssb.2220800164}{physica status solidi (b) 80~(1) (1977) K67--K70}.
\newblock \href{http://dx.doi.org/10.1002/pssb.2220800164}
  {\path{doi:10.1002/pssb.2220800164}}.

\bibitem{mijn.77}
P.~E. Mijnarends, The investigation of electron momentum distributions, in:
  B.~G. Williams (Ed.), Compton Scattering, McGraw-Hill, New-York, 1977, pp.
  323--345.

\bibitem{schw.04}
P.~Schwerdtfeger, Relativistic Electronic Structure Theory, Elsevier,
  Heidelberg, 2004.

\bibitem{borl.98}
L.~Borland, Ito-{L}angevin equations within generalized thermostatistics,
  Physics Letters A 245~(1-2) (1998) 67--72.

\bibitem{lutz.03}
E.~Lutz, Anomalous diffusion and {T}sallis statistics in an optical lattice,
  Phys. Rev. A 67 (2003) 051402.

\bibitem{lutz.04}
E.~Lutz, Power-law tail distributions and nonergodicity, Phys. Rev. Lett. 93
  (2004) 190602.

\bibitem{lutz.13}
E.~Lutz, F.~Renzoni, Beyond {B}oltzmann-{G}ibbs statistical mechanics in
  optical lattices, Nat Phys 9~(10) (2013) 615--619.

\bibitem{vi.pl.12}
C.~Vignat, A.~Plastino, A.~R. Plastino, J.~S. Dehesa,
  Quantum potentials with q-{G}aussian ground states,
  \href{http://www.sciencedirect.com/science/article/pii/S0378437111007692}{Physica A: Statistical Mechanics and its Applications 391~(4) (2012) 1068 -- 1073}.
\newblock \href{http://dx.doi.org/http://dx.doi.org/10.1016/j.physa.2011.09.031}
  {\path{doi:http://dx.doi.org/10.1016/j.physa.2011.09.031}}.

\end{thebibliography}

\end{document}